\documentclass{article}
\renewcommand{\baselinestretch}{1}
\usepackage{graphicx,spconf}
\usepackage{algorithm}
\usepackage{algpseudocode}%
\usepackage{enumerate}
\usepackage{amsthm}
\usepackage[small,compact]{titlesec}
\makeatletter
\def\th@newremark{\th@remark\thm@headfont{\bfseries}}
\makeatletter
\usepackage[cmex10]{amsmath}
 {  
    \newtheorem{proposition}{Proposition}
    
}
\theoremstyle{newremark}

\theoremstyle{newremark}
\newtheorem{rem}{Remark}
\theoremstyle{newremark}
\newtheorem{fact}{Fact}
\theoremstyle{newremark}
\newtheorem{problem}{Problem}
\theoremstyle{newremark}

\theoremstyle{newremark}

\theoremstyle{definition}

\theoremstyle{newremark}

\DeclareMathOperator*{\argmin}{arg\,min}
\usepackage{amssymb}

\title{\textsc{Set-Theoretic Learning for Detection in Cell-Less C-RAN Systems}}
\name{Daniyal Amir Awan$^{\star}$ \qquad Renato L.G. Cavalcante$^{\star \dagger}$ \qquad Zoran Utkovski $^{\dagger}$ \qquad Slawomir Stanczak$^{\star \dagger}$}
\address{$^{\star}$Technische Universitaet Berlin, Berlin, Germany\\
  $^{\dagger}$Fraunhofer Heinrich Hertz Institute, Berlin, Germany}

\begin{document}
\setlength{\abovedisplayskip}{4.0pt}
\setlength{\belowdisplayskip}{4.0pt}
\renewenvironment{enumerate}%
  {\begin{list}{\arabic{enumi}.}%
     {\topsep=0.025in\itemsep=0.025in\parsep=0.025in\partopsep=0.025in\usecounter{enumi}}%
   }{\end{list}}
%
\maketitle
\begin{abstract}
Cloud-radio access network (C-RAN) can enable cell-less operation by connecting distributed remote radio heads (RRHs) via fronthaul links to a powerful central unit. In conventional C-RAN, baseband signals are forwarded after quantization/compression to the central unit for centralized processing to keep the complexity of the RRHs low. However, the limited capacity of the fronthaul is thought to be a significant bottleneck in the ability of C-RAN to support large systems (e.g. massive machine-type communications (mMTC)). Therefore, in contrast to the conventional C-RAN, we propose a learning-based system in which the detection is performed locally at each RRH and only the likelihood information is conveyed to the CU. To this end, we develop a general set-theoretic learning method to estimate likelihood functions. The method can be used to extend existing detection methods to the C-RAN setting.
\end{abstract}
\section{Introduction}\label{sec:introduction}
Massive connectivity, especially in the context of massive machine-type communications (mMTC) and internet of things (IoT), is a cornerstone of future wireless networks. These next-generation systems (also known as fifth-generation (5G) systems) will comprise a large number of low-rate devices transmitting in the uplink \cite{Bockelmann2016}. Current cellular systems are not designed to deal with, among other things, a large signaling overhead (e.g. due to handovers) caused by an ever increasing number of devices in dense small-cell deployments \cite{Han2017}. As a result, the integration of novel network architectures with new data communication techniques has captured recent interest among wireless network operators and researchers alike. 
To this end, ``cell-less" systems \cite{Han2017,Ngo2017} have been recently proposed. These systems envision devices broadcasting to multiple transmission and reception points (TRPs), without having to associate with any TRP. 

Cloud-radio access network (C-RAN) is envisaged to be a key enabler of cell-less uplink because of its low cost and spectrum efficiency \cite{Tang2017,Checko2015}. In conventional C-RAN, joint baseband processing at centralized cloud processors or a central unit is performed on behalf of distributed TRPs called remote radio heads (RRHs).
			This migration of processing is made possible by deploying fronthaul links between RRHs and the central unit. Under the assumption of high-capacity fronthaul links, the cost reduction by using low-complexity RRHs is complemented by performance benefits emanating from joint detection/processing at the central unit \cite{Ngo2017}. The conventional C-RAN with the \textit{common public radio interface} specification prescribes simple scalar quantization for fronthaul links, but the performance of this approach degrades in the presence of stringent fronthaul capacity constraints \cite{Park2014}. To improve system performance in the case of low-capacity fronthaul, more sophisticated ``network aware" fronthaul compression has been proposed (see. e.g. \cite{Park2014}), where both the decompression of forwarded RRH signals and decoding/detection of user data takes place jointly at the central unit. These information-theoretic approaches study existence of coding schemes to achieve some sum-rate performance bounds. However, the results are based on asymptotic analysis and they assume coding over long blocks. Furthermore, these methods require some knowledge of the network (e.g. user channels and other network statistics) at the central unit. Therefore, these methods can be difficult to implement in practice in large networks. In contrast, we are concerned with ensuring reliable machine-type communication (in terms of bit error rate (BER)) at a fixed communication rate, and our learning-based approach results in a practical scheme that does not require any knowledge of user channels. The method can be used to extend existing detection mechanisms to the cell-less setting, and can be combined with any forward error correction (FEC)/coding scheme.
			
			In more detail, it has been shown that, in contrast to conventional C-RAN, it maybe advantageous to apply ``local'' pre-processing/detection at the RRHs followed by data fusion at the CU \cite{Utkovski2017}. Following this idea, we develop a learning-based ``detect and forward" scheme, whereby the likelihood ratios associated with local detection are combined at the CU to obtain the final estimate. We note that, whereas in the Bayesian detection techniques the likelihood information may come naturally, in non-Bayesian methods considered in this study, this is not the case. To this end, the main contribution of this study is a general set-theoretic low-complexity learning method to estimate probability density functions (pdfs), which we use to obtain a reliable approximation of likelihood functions. In this way, existing detection methods can be extended to the cell-less C-RAN setting. 
			We note that the recent study in \cite{Traganitis2017} proposes a similar practical and ``blind" approach to combining beliefs from multiple classifiers by estimating the so-called ``confusion matrix". This method, although demonstrated to work better than other state-of-art methods, is based on a heuristic solution to a non-convex optimization problem. The method proposed in this study, in contrast, solves a convex problem and the proposed algorithm provides convergence guarantee. Simulation results demonstrate the potential of this learning-based detection framework to overcome the fronthaul capacity limitation, and it delivers a significantly better detection performance than both of it's conventional counterpart and the method in \cite{Traganitis2017}.



\textit{\textbf{Notation}}: The sets of real numbers, non-negative integers, positive integers, and complex numbers are denoted by $\mathbb{R}$, $\mathbb{Z}_{\geq 0}$, $\mathbb{Z}_{>0}$, and $\mathbb{C}$, respectively.  
We define $\overline{N_{1},N_{2}}:=\left\{N_1,N_{1}+1,\ldots,N_2\right\}$, $N_1, N_2 \in \mathbb{Z}_{\geq 0}$ with $N_1\leq N_2$. 
Let $\mathcal{H}$ be a real Hilbert space with an inner product $\langle\cdot,\cdot\rangle$ and the induced norm $\left\|f\right\|^{2}_{\mathcal{H}}=\langle f,f \rangle$. For every $x\in\mathcal{H}$, the projection $P_{C}(x)$ onto a non-empty closed convex set $C\subset\mathcal{H}$ is the solution to the problem: $\inf_{y\in C}\frac{1}{2}\left\|x-y\right\|^{2}_{\mathcal{H}}$. A well-known result is that $P_{C}(x)$ always exists and is unique \cite[Th.~1, Ch.~3.12]{Luenberger1997}. 

\section{System Model}\label{sec:system_model}
We consider a C-RAN system \cite{Checko2015, Utkovski2017} consisting of $R$ RRHs each of which has $M$ antennas.  Each RRH is connected to a central unit by a capacity-limited and interference-free fronthaul link, whose capacity is bounded above by $B_{\text{p}}$ bits per packet. We consider the uplink where $K$ single-antenna devices broadcast their data to the RRHs \cite{Han2017,UCNC,Ngo2017}. The following techniques work on real vectors but they can be applied to the complex case by using the well-known bijection between an $M$-dimensional complex vector and $2M$-dimensional real vectors (see, e.g. \cite{AwanICC2018,Slavakis2009}). For clarity of presentation, we consider BPSK modulation in the following. The extension to higher modulation schemes is straight forward, and simulations in Section \ref{sec:results} are performed for QPSK modulation. 

The real-valued uplink received signal (sampled at a fixed symbol rate and assuming non-dispersive channels) at RRH $l\in\overline{1,R}$ is given by,
\begin{equation}\label{eqn:uplink_signal}
	\mathbf{r}^{l}:\mathbb{Z}_{\geq 0}\rightarrow \mathbb{R}^{2M}:t \mapsto \sum_{k=1}^{K}\sqrt{p_{k}}b_{k}(t)\mathbf{s}^{l}_{k}(t) + \mathbf{n}^{l}(t),
\end{equation} 
where $b_{k}(t)\in \{+1,-1\}$ and $p_{k}\in\mathbb{R}$ are, respectively, the BPSK symbol and the (fixed) transmit power of device $k \in \overline{1,K}$. The vectors $\mathbf{s}^{l}_{k}(t)\in\mathbb{R}^{2M}$ and $\mathbf{n}^{l}(t)\in\mathbb{R}^{2M}$ denote the channel signature of device $k$ and additive noise at RRH $l$, respectively. Note that the channel signature $\mathbf{s}^{l}_{k}(t)$ contains both the path-loss and small-scale fading that is assumed to have a Rayleigh distribution.

Many mobile communication systems perform channel estimation or learning of other parameters before the actual data communication takes place \cite{AwanICC2018,Du2017,Wen2016}. Under the assumption of Rayleigh block fading \cite{Marzetta1999}, learning (through training) and data communication is performed within each coherence block which is defined as a block of channel symbols over which the channel is assumed to be constant. We use $T_\text{c}$ to denote the length of the coherence block, and we assume that the first $T_\text{t}<T_\text{c}$ channel symbols are used for training. In the remaining time period of $T_\text{c}-T_\text{t}$, data communication can be performed provided that there exists a detection filter $f^{k}_{l}:\mathbb{R}^{2M} \rightarrow \mathbb{R}$ to detect the modulation symbol of device $k \in \overline{1,K}$ reliably. In the following, we omit the index $k$ since the same processing is applied to each device in parallel. 

\subsection{Learning-Based Detect-and-Forward Strategy}\label{sec:adaptive_filtering}
In conventional C-RAN, the received signal \eqref{eqn:uplink_signal} is simply quantized and forwarded to the central unit for centralized processing. We refer to this strategy as \textit{quantize-and-forward} (Q\&F). 
In contrast to this, the focus of this study is a learning-based \textit{detect-and-forward} (D\&F) approach which consists of following steps: 


\begin{enumerate}
\item During time period $T_{\text{t}}$, each RRH $l\in\overline{1,R}$ performs the training to learn a detection filter $f_l$ such that $(\forall t \in \mathbb{Z}_{\geq 0})$ $f_l(\mathbf{r}^l(t))= b(t) + \widetilde{n}(t)$, where $\widetilde{n}(t)$ is the residual interference and noise. Note that since $\mathbf{r}^l(t)$ is random, $f_l(\mathbf{r}^l(t))$ is also random.  
The training is performed using a training sequence $(\mathbf{r}^{l}(t),b(t))_{t\in\overline{0,T_\text{t}-1}}$. It is important to mention here that $f_l$ can be any appropriate detection method, and in the simulations we use the method in \cite{AwanICC2018} which serves as an example. Additionally, each RRH learns likelihood functions $\varphi^{l}(f_l(\mathbf{r}^l(t))|+1)=\mathbf{P}(+1|f_l(\mathbf{r}^l(t)))$ and $\varphi^{l}(f_l(\mathbf{r}^l(t))|-1)=\mathbf{P}(-1|f_l(\mathbf{r}^l(t)))$, where $\mathbf{P}(+1|f_l(\mathbf{r}^l(t)))$ and $\mathbf{P}(-1|f_l(\mathbf{r}^l(t)))$ are the posterior distributions.\footnote{We assume modulation symbols are equiprobable. Furthermore, the channel of each device to each RRH is assumed to be uncorrelated.} The approximation of likelihood functions (in Section \ref{sec:estimation_of_probability_density_functions}) is the main technical contribution of this study. 
\item During data communication, the RRH calculates two likelihood values $\mathcal{L}_{l}(+1;\mathbf{r}^{l}(t)):=\varphi^{l}(f_l(\mathbf{r}^{l}(t))|+1)$ and $\mathcal{L}_{l}(-1;\mathbf{r}^{l}(t)):=\varphi^{l}(f_l(\mathbf{r}^{l}(t))|-1)$. 
\item The central unit performs a maximum likelihood estimation of ${b}(t)$ given by\footnote{The log-likelihood ratios in \eqref{eqn:ml_decision} can be combined at the central unit by using various methods including consensus and optimal log-likelihood quantization approaches \cite{RenatoMeYamada2014,Rave2009}. These approaches are not the focus of the study and they are left for future work.} 
\begin{equation}\label{eqn:ml_decision}
    \hat{b}(t)=\text{sgn}\Bigg(\sum^{R}_{l=1}\log\frac{\mathcal{L}_{l}(+1;\mathbf{r}^{l}(t))}{\mathcal{L}_{l}(-1;\mathbf{r}^{l}(t))}\Bigg),\\
\end{equation}   
where $\text{sgn}(x)=+1$ if $x\geq 0$, otherwise $\text{sgn}(x)=-1$. 
\end{enumerate}


In the next section, we present a method to reliably estimate likelihood functions and omit the index $l$ because the same processing is applied at each RRH. 

\section{Estimation of Likelihood Functions}\label{sec:estimation_of_probability_density_functions} 
In this section, we present a general technique for obtaining a reliable approximation of a pdf given a sample set of independent and identically distributed (i.i.d) samples. We denote the pdf  of a random source $\mathbf{X}$ by $\varphi_\mathbf{X}$, and perform a ``set theoretic" approximation of $\varphi_\mathbf{X}$ by utilizing available prior knowledge. The prior knowledge includes general properties of pdfs and also knowledge derived from a given sample set $\mathcal{D}_{\mathbf{X}}:=\{x_1, x_2,\ldots, x_N\}$, which we assume  consists of i.i.d observations of $\mathbf{X}$. The sample set for likelihood functions can be generated by observing the response of the (trained) filter $f$ to the training sample set, after the training has been completed. As a particular example, let $\varphi_\mathbf{X}:=\varphi(f(\mathbf{r}^l(t))|+1)$ denote the likelihood function of the filter response given $b(t)=+1$, and recall that a training sequence $(\mathbf{r}(t),b(t))_{t\in\overline{0,T_\text{t}-1}}$ is known at the RRH at $t=T_\text{t}-1$ (see the D\&F process above). Then, we can extract a sample set $\mathcal{D}_{\mathbf{X}}:=\{f(\mathbf{r}(t))|b(t)=+1, t\in\overline{0,T_\text{t}-1}\}$ for $\varphi_\mathbf{X}$. The same applies to the case when $\varphi_\mathbf{X}:=\varphi(f(\mathbf{r}(t))|-1)$.

\subsection{Set Theoretic Approximation}\label{sec:best_approximation}
      We start by assuming that $\varphi_{\mathbf{X}} \in L^2(\mathbb{R})$, where $L^2(\mathbb{R})$ (henceforth denoted by $L^2$) is the Hilbert space of square (Lebesgue) integrable functions equipped with the inner product $(\forall f,g \in L^2)$ $\langle g,f \rangle_{L^2}:=\int_{\mathbb{R}}g(x)f(x)dx$ and the norm $\|f\|_{L^2}^{2}=\langle f,f \rangle_{L^2}<\infty$. 
			
			We base our method on that in \cite[Ch.~6.5]{Stark1998}, but in contrast to \cite[Ch.~6.5]{Stark1998}, we assume that $\varphi_{\mathbf{X}}$ belongs to a closed-subspace of $L^2$. In more detail, for fixed $N \in \mathbb{Z}_{\geq 0}$, $(\forall i\in\overline{1,N})$ $x_i \in \mathbb{R}$, and $\sigma > 0$, the space $\mathcal{G}:=\{\varphi\in L^2|\varphi=\sum^{N}_{i=1}w_{i}\kappa(\cdot,x_i), (\forall i \in\overline{1,N})w_i \in\mathbb{R}\}$, $(x \in \mathbb{R})$ $\kappa(x,x_i):=(1/\sqrt{2\pi\sigma^{2}})\exp\left(\frac{-|x-x_i|^{2}}{2\sigma^{2}}\right)$, is a closed subspace of $L^2$ \cite{Smola1998,Ohnish2017}. We equip $\mathcal{G}$ with the inner-product $\langle h,p\rangle_{\mathcal{G}}=\langle h,p \rangle_{L^2}$ and the norm $\|f\|_{\mathcal{G}}^{2}=\langle f,f \rangle_{\mathcal{G}}$ such that $\mathcal{G}$ is a Hilbert space. The reason for working with $\mathcal{G}$ is that the inner-products in $\mathcal{G}$ have closed-form solutions, which are computationally convenient for the algorithms presented below, and functions in $\mathcal{G}$ can well approximate continuous functions with compact support if $N$ is sufficiently large. Furthermore, during data communication, likelihood values $\varphi_\mathbf{X}(x)$ can be estimated by fast evaluations $\varphi^{\ast}(x)=\sum^{N}_{i=1}w_{i}\kappa(x,x_i)$, given a closed-form approximation $\varphi^{\ast}$ with weights $w_{i}$.
		
		 In light of the above, the objective now becomes to find a $\varphi^{\ast} \in \mathcal{G}$ that is in agreement with all the available information we have about $\varphi_{\mathbf{X}}$. More precisely, suppose that the prior information amounts to the fact that $\varphi_{\mathbf{X}}$ is a member of $Q$ closed-convex sets, i.e. $(\forall q\in\overline{1,Q})$ $\varphi_{\mathbf{X}}\in C_q \subset \mathcal{G}$. Then a reasonable approximation of $\varphi_{\mathbf{X}}$ is a solution to the \textit{set feasibility problem}: find $\varphi^{\ast}\in \mathcal{G}$ such that $\varphi^{\ast} \in \bigcap^{Q}_{q=1} C_q$. Set feasibility problems can be solved by a plethora of projection algorithms which are well-known for their simplicity \cite{Censor2012}. Moreover, we shall see in the following sections that some projection operations have low-complexity closed-forms in $\mathcal{G}$. 
		
		Before we proceed further, we describe some basic results pertaining to the subspace $\mathcal{G}$ to be utilized in the following sections. Denote by $\mathbf{G} \in \mathbb{R}_{\geq 0}^{N\times N}$ the positive semidefinite Gramm matrix with entries $(\forall i,j \in \overline{1,N})$ $[\mathbf{G}]_{i,j}:=\langle \kappa(\cdot,x_i),\kappa(\cdot,x_j)\rangle_{\mathcal{G}}$. 
The projection of $h$ onto $\mathcal{G}$ denoted by $P_{\mathcal{G}}(h)$ is given by $P_{\mathcal{G}}(h)=\sum^{N}_{i=1}\zeta_i(h) \kappa(\cdot,x_i)$; $(\forall i \in \overline{1,N})$ $\zeta_i(h)\in\mathbb{R}$ is the $i$th component of $\boldsymbol{\zeta}(h)$, where $\boldsymbol{\zeta}(h)$ is the solution to $\mathbf{G} \boldsymbol{\zeta}(h)=[\langle h,\kappa(\cdot,x_1)\rangle_{\mathcal{G}},\cdots,\langle h,\kappa(\cdot,x_N)\rangle_{\mathcal{G}}]^{\intercal}$ \cite[Ch.~6.9 , Ch.~3.6]{Luenberger1997}. 
		
		In the following, we show how to construct closed-convex sets (along with the corresponding projections) based on two sources of prior information: Sets $C_1,C_2,\ldots,C_Q$ are constructed (in Section \ref{sec:sample_sets}) based on the \textit{sample set} $\mathcal{D}_{\mathbf{X}}$ specific to $\varphi_{\mathbf{X}}$, while $C_{Q+1}$ and $C_{Q+2}$ (in Section \ref{sec:normalization} and Section \ref{sec:positivity} resp.) are constructed based on \textit{necessary conditions} for pdfs. The projections are then utilized by the iterative algorithm in Section \ref{sec:proj_algorithm} to solve: 
		\begin{problem}
		Find a  $\varphi_\mathbf{X}$ such that $\varphi_\mathbf{X}\in \bigcap_{q \in \overline{1,Q+2}} C_{q}$, under the assumption that $\bigcap_{q \in \overline{1,Q+2}} C_{q} \neq \emptyset$, where sets $({q \in \overline{1,Q+2}})$ $C_{q}$ are defined below.
		\label{problem:one}
		\end{problem}
		Note that the proposed algorithm can also deal with the case when $\bigcap_{q \in \overline{1,Q+2}} C_{q}=\emptyset$ (see Remark 1). In the following, we denote by $\varphi_{(n)}$ the $n$th iteration of the algorithm in Section \ref{sec:proj_algorithm}.

\subsubsection{Convex Sets Based on the Sample Set}\label{sec:sample_sets}
     Consider the event $\{a_q \leq \mathbf{X} \leq b_q\}$ and suppose that the probability of this event $\text{Pr}[a_q \leq \mathbf{X}\leq b_q]$ is unknown. Given a sample set $\mathcal{D}_{\mathbf{X}}:=\{x_1, x_2,\ldots, x_N\}$, we can divide the range of values in $\mathcal{D}_{\mathbf{X}}$ in disjoint intervals $(q\in\overline{1,Q})$ $[a_q, b_q]$, where $Q$ is a design parameter. Let $\overline{p}_q:=\text{Pr}[a_q \leq \mathbf{X}\leq b_q]$ and note that since the interval $[a_q, b_q]$ is obtained from $\mathcal{D}_{\mathbf{X}}$, $\overline{p}_q$ as a function of $[a_q, b_q]$ is a random variable. We follow the approach in \cite[Ch.~6.5]{Stark1998} to calculate the $95\%$ confidence interval $\mathcal{P}_q:=[P^{\text{L}}_q,P^{\text{H}}_q]$ for each $\overline{p}_q$ such that $(\forall q\in\overline{1,Q})$ $\text{Pr}[P^{\text{L}}_q \leq \overline{p}_q \leq P^{\text{H}}_q]\approx 0.95$. These calculations are computationally inexpensive. We omit the details here due to space limitation. 
		Since $\varphi_\mathbf{X}$ is the pdf of $\mathbf{X}$, it must be a member of every $C_{q}$ given by $C_{q}:=\{\varphi \in \mathcal{G}|\text{Pr}[a_q \leq \mathbf{X}\leq b_q]=\int_{a_q}^{b_q}\varphi(x)dx \in \mathcal{P}_q\}$. The integral $\int_{a_q}^{b_q}\varphi(x)dx$ can be written as the inner-product\footnote{All inner-products (integrals) involved in the projections have well-known closed forms which we omit to save space and maintain clarity of text.} $\langle P_{\mathcal{G}}(\mathbf{1}^{q}),\varphi\rangle_{\mathcal{G}}=\int_{-\infty}^{\infty}\mathbf{1}^{q}(x)\varphi(x)dx$; $\mathbf{1}^{q}(x)=1$ if $x\in[a_q,b_q]$, otherwise $\mathbf{1}^{q}(x)=0$. 
		
		The projection $P_{C_{q}}(\varphi_{(n)})$ onto the closed-convex set $C_{q}$ is given as
	\[ 
	P_{C_{q}}(\varphi_{(n)})=\Bigg\{\begin{tabular}{cc}
	$\varphi_{(n)}-\frac{q^q-P^{\text{H}}_q}{\|P_{\mathcal{G}}(\mathbf{1}^{q})\|^{2}_{\mathcal{G}}}P_{\mathcal{G}}(\mathbf{1}^{q})$,& if $q^q-P^{\text{H}}_q>0$\\
	$\varphi_{(n)}-\frac{q^q-P^{\text{L}}_q}{\|P_{\mathcal{G}}(\mathbf{1}^{q})\|^{2}_{\mathcal{G}}}P_{\mathcal{G}}(\mathbf{1}^{q})$,&  if $q^q-P^{\text{L}}_q<0$\\
	$\varphi_{(n)}$, &\text{otherwise}.
		\end{tabular}
	\]
	where $q^q:=\langle P_{\mathcal{G}}(\mathbf{1}^{q}),\varphi_{(n)}\rangle_{\mathcal{G}}$.


\subsubsection{Convex Sets Based on the Normalization Property}\label{sec:normalization}
A necessary condition is that $(\forall x \in \mathbb{S})$ $\int_{\mathbb{S}}\varphi_\mathbf{X}(x)dx=1$; $\mathbb{S}$ is the support of $\varphi_\mathbf{X}$ which we assume to be bounded. This implies that $\int_{-\infty}^{\infty}\mathbf{1}^{\mathbb{S}}(x)\varphi(x)dx=1$; $\mathbf{1}^{\mathbb{S}}(x)=1$ if $x \in \mathbb{S}$, otherwise $\mathbf{1}^{\mathbb{S}}(x)=0$. The projection $P_{C_{Q+1}}(\varphi_{(n)})$ onto the closed-convex set $C_{Q+1}=\{\varphi\in\mathcal{G}|\langle P_{\mathcal{G}}(\mathbf{1}^{\mathbb{S}}),\varphi\rangle_{\mathcal{G}}=\int_{-\infty}^{\infty}\mathbf{1}^{\mathbb{S}}\varphi(x)dx=1\}$, is given by
\begin{equation} 
P_{C_{Q+1}}(\varphi_{(n)})= \varphi_{(n)}-\frac{\langle P_{\mathcal{G}}(\mathbf{1}^{\mathbb{S}}),\varphi_{(n)}\rangle_{\mathcal{G}} -1}{\|P_{\mathcal{G}}(\mathbf{1}^{\mathbb{S}})\|^{2}_{\mathcal{G}}}P_{\mathcal{G}}(\mathbf{1}^{\mathbb{S}}).\nonumber
\end{equation}
		
	\subsubsection{Convex Sets Based on the Non-negativity Property}\label{sec:positivity}
A necessary condition is that $(\forall x \in \mathbb{S})$ $\varphi_\mathbf{X}(x) \geq 0$. Let $\varphi_{(n)}=\sum^{N}_{i=1}v_{i}\kappa(\cdot,x_i)$ and $\mathbf{v}=[v_{1},v_{2},\cdots,v_{N}]^{\intercal}$. Then, a sufficient condition for non-negativity of $\varphi$ is that $(\forall i \in\overline{1,N})$ $v_i \geq 0$. Ensuring this condition entails projection onto the closed-convex cone $C_{Q+2}:=\{\varphi\in\mathcal{G}|\varphi=\sum^{N}_{i=1}w_{i}\kappa(\cdot,x_i), (\forall i \in\overline{1,N}) w_i \geq 0 \}$.
Due to space limitation, we omit the proof of the following assertion:
\begin{proposition}
 The projection $P_{C_{Q+2}}(\varphi_{(n)})$ is given as $P_{C_{Q+2}}(\varphi_{(n)})=\sum^{N}_{i=1}w_{i}\kappa(\cdot,x_i)$; $(i \in \overline{1,N})$ $w_{i}$ is the $i$th component of 
$\mathbf{w}^{\ast}\in \arg \min_{\mathbf{w}\geq0} \frac{1}{2}\mathbf{w}^{\intercal}\mathbf{G}\mathbf{w}-\mathbf{w}^{\intercal}\mathbf{G}\mathbf{v}$.
\label{prop:one}
\end{proposition}

Note that the above quadratic program (QP) can be solved by any standard convex solver.	
		
	
\subsubsection{Projection Algorithm}\label{sec:proj_algorithm}
We use the following parallel projection algorithm to solve Problem 1.  
\begin{fact}[\textbf{Parallel Projection Algorithm}] \cite[Corollary~2.10-1]{Stark1998}.
For every choice of $\varphi_{(0)} \in \mathcal{G}$ and every choice of $(q\in\overline{1,Q+2})$ $\beta_{q}>0$ such that $\sum^{Q+2}_{q=1}\beta_{q}=1$, the sequence $\varphi_{(n)}$	generated by
   \begin{equation}
                 \varphi_{(n+1)}=\sum^{Q+2}_{q=1} \beta_{q} P_{C_{q}}(\varphi_{(n)})	
    \label{eqn:simultaneous_projection_algorithm}															
   \end{equation}
converges to $\varphi^{\ast}\in \bigcap_{q \in \overline{1,Q+2}} C_{q} \subset \mathcal{G} \subset L^2$. 
\label{def:best_definition} 
\end{fact}  
%
  \begin{rem} [\textbf{Empty Intersection and} $\beta_{q}$]   
	The algorithm in Fact $\ref{def:best_definition}$ guarantees convergence to a point that minimizes the weighted sum of minimum distances from sets $C_q$, i.e. $\varphi^{\ast}\in \argmin \phi(\varphi)$, $\phi(\varphi):=\sum^{Q+2}_{q=1}\beta_{q}\|\varphi-P_{C_{q}}(\varphi)\|^{2}_\mathcal{G}$. Since the design parameters $\beta_{q}$ assign priorities to sets $C_q$, it is intuitive to set $(\forall q \in \overline{1,Q})$ $\beta_{Q+2}=\beta_{Q+1}>\beta_q$ to keep $\varphi^{\ast}$ close to important sets $C_{Q+1}$ and $C_{Q+2}$ in case $\bigcap_{q \in \overline{1,Q+2}} C_{q} =\emptyset$.
	\end{rem}
 \section{Simulation and Conclusion}\label{sec:results}
      In this section, we compare the performance of D\&F strategy with centralized Q\&F strategy for QPSK modulation for limited fronthaul capacity and also with that of the method in \cite{Traganitis2017} that can also be applied to our problem. To perform filtering at each RRH, we use the learning-based method in \cite{AwanICC2018} that has been shown to out perform the conventional \textit{MMSE-SIC} based systems. Following the approach of non-orthogonal multiple access systems, devices are assumed to be allocated to clusters that are assigned disjoint resource blocks (RBs) of the system spectrum (no inter-cluster interference).
			In the simulation, we only consider a single cluster of devices but the same processing is applied to each cluster in parallel. 
			The device SNRs $(k \in \overline{1,K})$ $\gamma_k$ at each RRH are chosen independently at random from the set $\{-3\,\text{dB},-2\,\text{dB},\cdots, 9\,\text{dB}, 10\,\text{dB}\}$. We observed that for SNR values in this range, device have a strong enough signal at the receiver to be detected. To obtain robust statistics, we performed $10000$ experiments with different $\gamma_k$ values (chosen independently at random) and Rayleigh channels to obtain average (Gray-coded) BER.

		To perform training for D\&F, we use Algorithm $1$ in \cite{AwanICC2018}. After the training phase, we use the algorithm in Fact \ref{def:best_definition} to estimate likelihood functions. The values in the sample set $\mathcal{D}_{\mathbf{X}}=\{x_1,x_2,\cdots,x_N\}$ are used for the parameters $x_i$ for funtions $\kappa(x,x_i):=(1/\sqrt{2\pi\sigma^2})\exp\left(\frac{-|x-x_i|^{2}}{2\sigma^{2}}\right)$ (see Section \ref{sec:best_approximation}), whereas the width $\sigma$ is chosen according to the \textit{Silverman's} rule of thumb $\sigma=1.06\widehat{\sigma}N^{-1/5}$; $\widehat{\sigma}$ is the standard deviation of the samples. Furthermore, we used $Q=10$ intervals in Section \ref{sec:sample_sets} and ran $20$ iterations of the algorithm in Fact \ref{def:best_definition}. We observed a good performance for these heuristics. To observe the effect of the limited-capacity fronthaul, the obtained functions are then quantized by using the \textit{Max-Llyod} algorithm for quantization bits satisfying $B_{\text{q}}\leq 4$. For quantization bits satisfying $B_{\text{q}}>4$, we use uniform quantization. The maximum likelihood decision is performed at the central unit by combining likelihood ratios associated with the local detection by each RRH (see Section \ref{sec:adaptive_filtering}). 
For Q\&F, we first collect the training data at each RRH. We then estimate and quantize the pdfs of received vectors by the same process as in the D\&F case. At the central unit, the quantized vectors $\tilde{\mathbf{r}}^{l}(t):=\mathcal{Q}_{l}(\mathbf{r}^{l}(t))$ obtained from each RRH $l \in \overline{1,R}$ are stacked to obtain a vector $\mathbf{r}\in\mathbb{R}^{2MR}$. The centralized learning is then performed at the central unit by using Algorithm $1$ in \cite{AwanICC2018} with the quantized training data obtained from the RRHs. The centralized detection is performed using a filter $f:\mathbb{R}^{2RM}\rightarrow\mathbb{R}$.

\begin{figure}[t]
 \centering
 \includegraphics[width=0.5\textwidth]{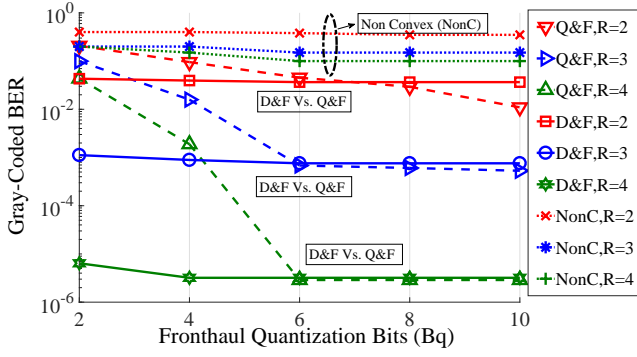}
  \caption{Comparison between the non-convex method (NonC) \cite{Traganitis2017}, D\&F, and Q\&F: $M=3$, $T_{\text{t}}=100$, $K=6$, $R\in\{2,3,4\}$}
\label{fig:spawc_fig}
\end{figure}

		Figure \ref{fig:spawc_fig} shows the average (Gray-coded) bit error rate (BER) for QPSK modulation for increasing values of fronthaul packet lengths which result in quantization bits $B_\text{q}=B_\text{p}/2K$ per user in the D\&F case, and $B_\text{q}=B_\text{p}/2M$ per receive vector component in the Q\&F case. We compare D\&F (in solid-lines) and Q\&F (in dashed-lines) forwarding strategies as described in Section \ref{sec:system_model} for different values of number of RRHs $R$. The D\&F strategy developed in this study clearly outperforms the Q\&F one for a low fronthaul capacity. On the other hand, Q\&F is more suited to situations with a large fronthaul capacity. Note that, intuitively, our method competes well with Q\&F because the likelihood approximation is sufficiently reliable, which is not the case with the learning framework (NonC, top 3 graphs) of \cite{Traganitis2017}. The reason is mainly the lack of sufficient training ($T_{\text{t}}=100$) to acquire statistics, but we conjecture that the performance also suffers from the lack of convexity of the optimization problem.   
		

        
To conclude, we proposed a learning-based ``detect-and-forward" scheme for cell-less C-RAN with low-capacity fronthaul. To this end, we presented a set-theoretic learning method to estimate likelihood functions that can be used to extend existing detection schemes to the cell-less setting. Simulation shows that our method outperforms both the conventional ``quantize and forward" method and a recently proposed comparable scheme for a low fronthaul capacity. 

\renewcommand{\baselinestretch}{1}
\bibliographystyle{IEEEtran}
\bibliography{library}

\end{document}